\documentclass[prl,twocolumn,showpacs,10pt]{revtex4}
\pdfoutput=1
\usepackage{graphicx}

\begin{document}
\title{Single Particle Tunneling in Strongly Driven Double Well Potentials}

\author{E.~Kierig$^1$}
\author{U.~Schnorrberger,$^1$ A.~Schietinger,$^1$ J.~Tomkovic,$^{1,2}$ and
M.K.~Oberthaler$^1$}

\affiliation{$^1$Kirchhoff-Institut f\"ur Physik, Universit{\"a}t
Heidelberg, Im Neuenheimer Feld 227, 69120 Heidelberg, Germany}
\homepage{www.kip.uni-heidelberg.de/matterwaveoptics}
\affiliation{$^2$Institut f\"ur Theoretische Physik, Technische
Universität Dresden, 01062 Dresden, Germany }


\date{\today}

\begin{abstract}
We report on the first direct observation of coherent control of
single particle tunneling in a strongly driven double well
potential. In our setup atoms propagate in a periodic arrangement of
double wells allowing the full control of the driving parameters
such as frequency, amplitude and even the space-time symmetry. Our
experimental findings are in quantitative agreement with the
predictions of the corresponding Floquet theory and are also
compared to the predictions of a simple two mode model. Our
experiments reveal directly the critical dependence of coherent
destruction of tunneling on the generalized parity symmetry.
\end{abstract}




 \maketitle

Strongly driven quantum systems are intensively studied as they
appear in many different contexts in physics and chemistry
\cite{driven quantum systems}. Here we are interested in the
situation of coherent control of tunneling due to strong driving. It
has been pointed out in 1991 \cite{Haenggi91} that for a specific
driving of a double well system the tunneling dynamics can be
brought to a complete standstill known as coherent destruction of
tunneling (CDT).

Very recently this effect in a double well situation has been
visualized in specially designed optical wave guides
\cite{optics_cdt} and a closely related effect in a periodic
potential has been observed in the nonlinear dynamics of a
Bose-Einstein condensate \cite{morsch}. Here we report on the first
demonstration of this effect for single particle tunneling in a
double well potential. Since our system allows in a straight forward
manner to manipulate the spatial as well as the temporal symmetry of
the driving we can demonstrate experimentally for the first time the
predicted dependence on the generalized parity symmetry
$P:x\rightarrow-x, t\rightarrow t+\frac{T}{2}$ of the situation
where $T$ represents the period of the driving
\cite{Milena_Grifoni}. With the same setup also the regime of
accelerated tunneling can be realized by choosing the proper driving
frequency \cite{acceleration_of_tunneling}.

In the following we present our systematic investigations on the
control of single particle tunneling dynamics in a double well
potential. We find quantitative agreement of the observed increase
as well as decrease of the tunneling rate with the theoretical
prediction of a Floquet analysis \cite{Milena_Grifoni}. The direct
connection of the observed slowing down with the effect of coherent
destruction of tunneling has been confirmed observing its strong
dependence on the space-time symmetry of the driving force.
\begin{figure}[h]
\begin{center}
\includegraphics[width=0.8\linewidth]{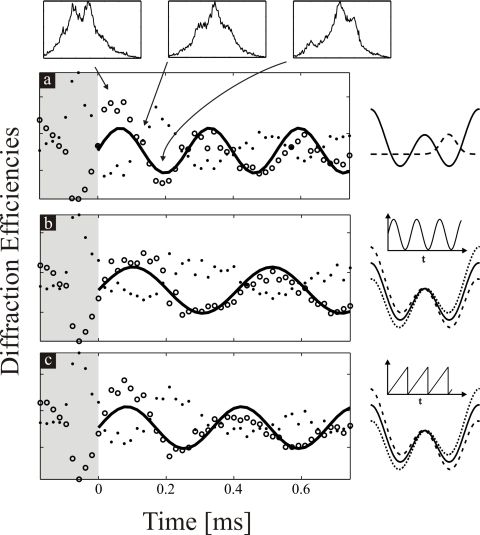}
\caption{\label{fig1} Coherent control of single particle tunneling.
(a) A single atom is realized on the right side of the double well
potential and the subsequent tunneling dynamics can be directly
observed in momentum space. Since a periodic arrangement of double
well potentials is used the momentum distribution is given by a
diffraction pattern as shown in the insets. (b) Adding symmetric
driving leads to the predicted slowing of the dynamics. (c) Breaking
the generalized parity symmetry by applying a sawtooth driving force
destroys this effect. The gray shaded area indicates the preparation
phase. The solid lines are sinusoidal fits to the data.}
\end{center}
\end{figure}

The qualitative behavior of the tunneling dynamics in momentum space
is shown in fig.~\ref{fig1}. In the experiment few periodically
arranged double wells are coherently populated and thus the momentum
distribution is given by the corresponding diffraction pattern as
shown in the insets. The tunneling dynamics directly shows up in the
dynamics of the diffraction efficiencies. In fig.~\ref{fig1}(a) the
tunneling of a single atom in an unperturbed double well potential
is depicted. In fig.~\ref{fig1}(b) one can clearly see that the
dynamics is slowed down if symmetric driving is applied. Breaking
the temporal symmetry of this driving i.e. sawtooth shape, the
tunneling rate approaches the undriven tunneling rate (see
fig.~\ref{fig1}(c)). This is qualitatively expected from a simple
symmetry argument and as we will show later it is also in
quantitative agreement with theory.

\begin{figure}[h]
\begin{center}
\includegraphics[width=0.9\linewidth]{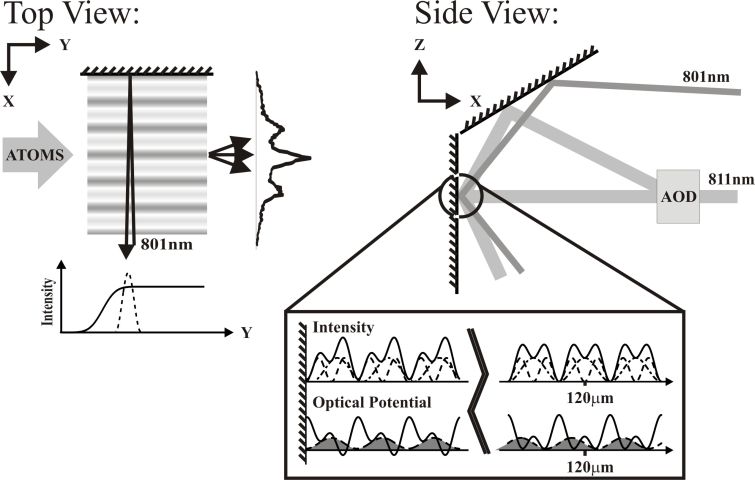}
\caption{\label{fig2} Schematics of the experimental setup. The well
collimated atomic beam impinges on periodic light shift potentials,
which adiabatically increase along the propagation direction. The
interaction time is adjusted by changing the extension of the light
field in y-direction. The momentum distribution is detected in the
far field by a multi-channel plate. The used optical setup with an
acousto optical deflector is depicted on the right hand side. In the
inset the change of symmetry due to residual phase shifts at the
mirror surface is shown.}
\end{center}
\end{figure}

Before we will discuss our systematic investigations and the
corresponding Floquet description we shortly discuss the
experimental setup which is schematically depicted in
fig.~\ref{fig2}. Since we are dealing with one dimensional physics a
traditional well collimated (FWHM $<$180$\mu$rad) atomic beam is
perfectly suited as a particle source and the dynamics of interest
happens only in the transverse direction. We have chosen metastable
argon since it allows for a spatially resolved single particle
detection using a multi-channel plate. Additionally for metastable
argon imaginary optical potentials \cite{imaginary_potential} can be
realized using resonant light at $801$nm. This transition quenches
the metastable state to the ground state which is not detectable
with a channel plate and thus can be employed for the preparation of
a particle on one side of the barrier.

The periodic arrangement of double wells is achieved employing
standard optical light shift potentials (see fig.~\ref{fig2}). By
retro reflecting two far red-detuned light beams
($\lambda=811.775$nm) realized with an acousto optical deflector
(AOD) under $0^{\circ}$ and $60^{\circ}$ incidence standing light
waves are formed with periodicities of $\lambda/2$ and $\lambda$
respectively in x-direction. Using a gold mirror one has to take
into account that the residual absorption leads to a phase shift
between the two light waves of $\phi=0.4$ rad for light polarized
perpendicular to the plane of incidence resulting in an arrangement
of asymmetric double wells. In our experiment we compensate this
phase shift i.e. asymmetry, by adjusting the reflection angle to
$59.97^\circ$ (by changing the AOD frequency) leading to a different
spatial periodicity such that the phase is compensated at a distance
of $120\mu$m from the mirror surface (see fig.~\ref{fig2}(c)). The
imaginary potential for the preparation of the atoms in one well is
implemented by retro reflection of $\lambda=801.702$nm light on the
same mirror but under an angle such that at a distance of $120\mu $m
the phase shift relative to the double wells is about $\pi /4$
leading to the preparation of the atom in one well.

The light intensity profile along the atomic beam is adjusted with
specially designed optical gray filters such that the motion of the
atoms follows adiabatically the light shift potentials. In the case
of perpendicular incidence of the atoms onto the standing light wave
the symmetric ground state of the double well potential is
populated. This critical alignment is achieved by standard Bragg
scattering. The potential heights are calibrated by either preparing
non adiabatically and observing the oscillation frequency
($\lambda/2$ periodicity) or preparing a wave packet such that it
oscillates around the potential minimum ($\lambda$ periodicity).

 The driving is realized by tuning the relative
phase between the two periodic potentials. This is experimentally
implemented by periodically changing the driving frequency of the
AOD i.e.~the diffraction angle. Since this is under full
experimental control it offers a great freedom in choosing frequency
as well as the space-time symmetry of the driving force.

In order to make a quantitative comparison of the experimental
results with theoretical predictions over the whole range of
experimentally accessible parameters we employ the standard Floquet
formalism. Our system is described by the Hamiltonian
\begin{equation}
H = \frac{p^{2}}{2m} + V_{1} \cos^{2}(kx) + V_{2}\cos^{2}(kx
\cos(\frac{\pi}{3}+\epsilon f(t)))
\end{equation}
with $V_{1,2}$ being the amplitudes of the two potentials forming
the double well structure and $\epsilon f(t)$ the deviation from the
incidence angle of $60^\circ$. In the limit of small $\epsilon$ this
leads to a driving potential of the form $V_d= S \sin(k \tilde x)
f(t)$ with $\tilde x=0$ at the symmetry of the double well
potential. $S$ is the amplitude of the driving and $f(t)$ describes
the time dependence of the driving force with the characteristic
driving frequency $\omega_d=2\pi/T$. As this Hamiltonian is time
periodic we can introduce the Hermitian Operator
$\mathcal{H}(x,t)=H(x,t)-i\hbar \frac{\partial}{\partial t}$ and
according to the Floquet theorem make a plane wave ansatz for the
state vector $|\Psi(t)_{\alpha}\rangle =
\exp(-i\epsilon_{\alpha}t/\hbar)|\Phi_{\alpha}(t)\rangle $ where
$|\Phi_{\alpha}(t)\rangle = |\Phi_{\alpha}(t+T)\rangle$. In doing so
we reduce the problem to solving the eigenvalue equation for
quasienergies $\epsilon_{\alpha}$
\begin{equation}
\mathcal{H}\Phi_{\alpha}(x,t)=\epsilon_{\alpha}\Phi_{\alpha}(x,t)
\end{equation}
which can be easily done numerically. For the time periodic function
$\Phi_{\alpha}$ we can make a Fourier expansion and choose the
eigenstates  of the unperturbed double well potential as an
orthogonal basis. For the results shown in the following we have
taken into account the 15 lowest energy eigenstates.

\begin{figure}[ht]
\begin{center}
\includegraphics[width=0.7\linewidth]{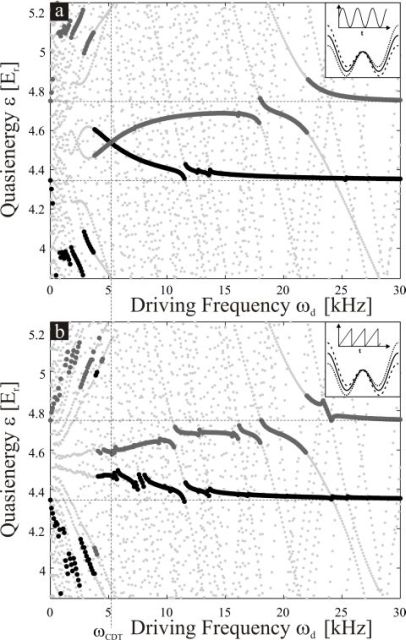}
\caption{\label{fig3} The Floquet state quasi energies for the
experimentally realized parameters. The initial preparation of a
particle localized on one side leads to the population of mainly two
Floquet states and the corresponding eigen energies are indicated as
dark-gray and black dots. The tunneling dynamics is given by the
quasi energy difference. (a) In the case of symmetric driving the
Floquet states cross for at $\omega_{\textrm{\tiny{CDT}}}$ leading
to coherent destruction of tunneling. (b) Breaking the temporal
symmetry as indicated in the inset introduces an anti-crossing.
Clearly for high driving frequencies independent on the driving
symmetry the tunneling splitting increases and thus faster tunneling
is expected.}
\end{center}
\end{figure}

In fig.~\ref{fig3} the results for the quasienergies are shown for
sinusoidal and sawtooth driving. The corresponding eigenenergies of
the eigenstates which are maximally populated due to our initial
condition of a particle localized on one side of the barrier are
indicated with black and dark-gray points. An exact crossing of the
relevant quasieinergies implies that the slow dynamics comes to a
complete stand still i.e.~CDT, indicated by the dashed vertical
line. This happens if the eigenstates of the crossing quasienergies
belong to different parity classes of the generalized parity
$P:x\rightarrow-x, t\rightarrow t+\frac{T}{2}$. Clearly in the case
of broken symmetry (fig.~\ref{fig3}(b)) no crossing exists.
Furthermore it becomes clear that the tunneling rate will increase
as the symmetry is broken. For high driving frequencies independent
of the symmetry the theory predicts an acceleration of the tunneling
in comparison to the undriven case. The dashed horizontal lines
indicate the eigenenergies of unperturbed double well.

\begin{figure}[ht]
\begin{center}
\includegraphics[width=0.65\linewidth]{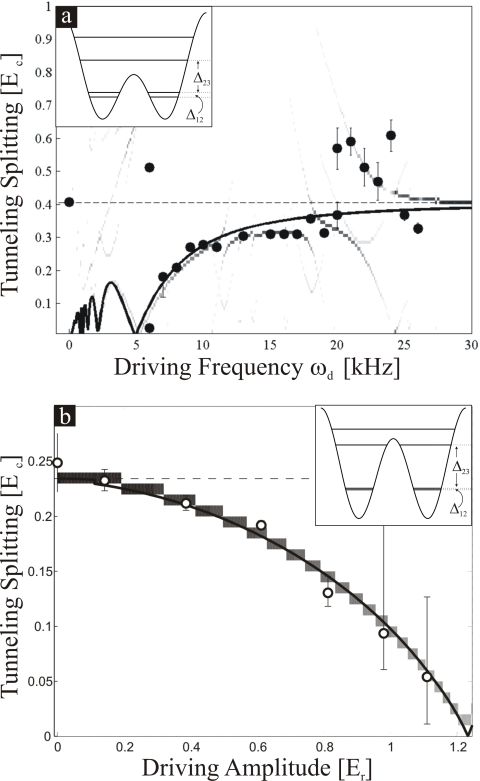}
\caption{\label{fig4}  Systematic study of the tunneling dynamics
with symmetric driving. (a) The graph shows the driving frequency
dependence of the tunneling rate. If the driving frequency is close
to the resonance to the third excited state the driving leads to an
acceleration of the tunneling. The experimental results (solid dots)
are explained by the full Floquet analysis without free parameter
over the whole range. The solid line represents the prediction
within the two mode approximation. (b) In this graph we compare our
experimental data obtained for fixed driving frequency but changing
the driving amplitude with the theoretical predictions. For this
specific experimental parameters also the two mode approximation
explains quantitatively the observations.}
\end{center}
\end{figure}

The results of our systematic investigations of ac-control of
coherent tunneling with symmetric driving are summarized in
fig.~\ref{fig4}. Plot (a) shows the tunneling splitting deduced from
the observed dynamics as a function of driving frequency as solid
dots ($S=0.88 E_{r}$ with $E_r=h^2/2m\lambda^2$ for
$\lambda=811$nm). The dashed line indicates the tunneling splitting
for the unperturbed double well potential. Since for a given driving
frequency more than two quasienergy differences are relevant we have
chosen a gray shading representing the weight given by the
population of the corresponding eigenstates due to the initial
condition i.e.~the probability to find this tunneling frequency. It
has to be noted that in the theoretical prediction no free parameter
is used. Thus we have very good quantitative agreement between
theory and experiment. Furthermore the results show that driving
allows for the full control of the tunneling dynamics
i.e.~acceleration as well as suppression.

For completeness we have added the analytical prediction within the
two mode approximation for the effective tunneling splitting
$\Delta_{\textrm{\tiny{eff}}} = J_{0}(\frac{Sx_{12}}{\hbar \omega})
\Delta_{12}$ where $J_0$ represents the zeroth-order Bessel
function, $x_{12} = \langle\varphi_{1}|\sin(kx)|\varphi_{2}\rangle$
is the transition dipole matrix element and $\Delta_{12}$ is the
tunneling splitting of the unperturbed system \cite{Milena_Grifoni}.
It is clear that the two mode approximation captures the CDT very
well. The deviation in respect to the Floquet analysis comes from
the fact that we are not deep in two mode regime as can be seen in
the inset ($V_{1}=6.25E_{r}, V_{2}=5.40E_{r}$). The increase of
tunneling rate for very high driving frequencies is due to the
resonance with the second excited state in the potential well. At
this resonance the tunneling dynamics follows from the interplay
between three Floquet states and thus is very similar to the physics
of chaos assisted tunneling \cite{chaos_assisted_tunneling}.

The dependence of the tunneling splitting as a function of driving
amplitude for fixed driving frequency $\omega_d = 6 $kHz is shown in
fig.~\ref{fig4}(b). There we compare our experimental results with
the Floquet and two mode theory. It is important to note that the
potential parameters ($V_{1}=8.27E_{r}, V_{2}=2.68E_{r}$) for these
experiments are deeper in the two mode regime (see inset in
fig.~\ref{fig4})and thus the two mode approximation fits perfectly
with the Floquet theory. Clearly the splitting gets smaller i.e.~the
dynamics gets slower as the amplitude is increased. Also here we get
quantitative agreement between theory and experiment without free
parameter.

\begin{figure}
\begin{center}
\vspace{0.2cm}
\includegraphics[width=0.7\linewidth]{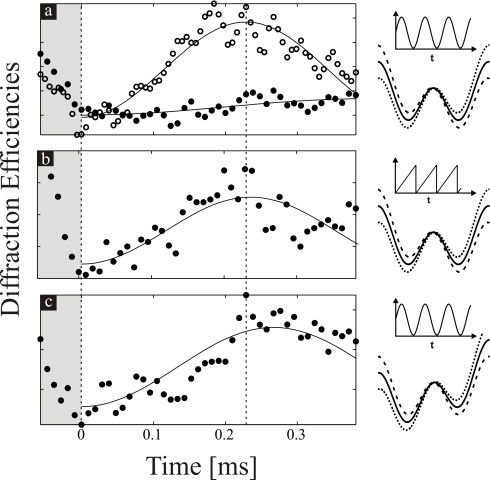}
\caption{\label{fig5}Tunneling dynamics for different driving
symmetries. (a) The open circles represent the tunneling dynamics in
a stationary potential while the solid dots reveal the slowing down
of the dynamics due to symmetric driving. (b) Applying a sawtooth
temporal driving to a spatially symmetric double well leads to
significant tunneling during the interaction time. (c) Breaking the
spatial symmetry i.e.~asymmetric double well, but employing
symmetric sinusoidal temporal driving also leads to the expected
increase of the tunneling rate.}
\end{center}
\end{figure}

In order to verify that CDT is indeed observed we demonstrate the
critical dependence of the slowing down of the tunneling on the
generalized parity symmetry of the driving $P:~x\rightarrow-x,
t\rightarrow t+\frac{T}{2}$. This has been implemented
experimentally in two different manners namely breaking the temporal
symmetry for a spatially symmetric double well (see
fig.~\ref{fig5}b) and breaking the spatial symmetry i.e. asymmetric
double well, with symmetric temporal driving (see fig.~\ref{fig5}c).
Clearly the tunneling dynamics is faster if the symmetry of the
driving is broken.

In this paper we have experimentally demonstrated the versatility of
strong driving as a new tool to modify the coherent tunneling
dynamics. In the experiment we have investigated the tunneling of a
single particle in a double well potential for different driving
situations. With that we clearly demonstrate experimentally the
critical dependence of coherent destruction of tunneling on the
underlying symmetry of the driving. We find excellent quantitative
agreement between experiment and theory. The realization of a
periodic potential with perfectly controllable parameters such as
symmetry of the unit cell and driving makes it a general model
system for studying strongly driven systems in the quantum regime
with the potential for preparation of complex quantum states in many
particle systems \cite{weis} but also extendable to the regime of
chaotic motion present in hamiltonian ratchets
\cite{Rachets_papers}.

\begin{acknowledgments}
We wish to thank Peter H\"anggi for the discussions initiating this
experiment. For the experimental contributions during the building
up phase and carefully reading the manuscript we thank Ralf
St\"utzle and Ramona Ettig. For support on the Floquet state
analysis we would like to thank Waltraud Wustmann and Roland
Ketzmerick. This work was financially supported by Landesstiftung
Baden-W\"urttemberg Atomics and the University of Heidelberg.
\end{acknowledgments}


\begin{thebibliography}{99}
\bibitem{driven quantum systems} W.~Domcke, P.~H\"anggi and D.~Tannor, eds., Special Issue: Chemical Physics, Vol. 217, No. 2-3: 117-416 (1997). 
\bibitem{Haenggi91} F.~Grossmann, T.~Dittrich, P.~Jung, and P.~H\"{a}nggi, Phys.~Rev.~Lett.~\textbf{67}, 516 (1991).
\bibitem{optics_cdt} G.~Della Valle, et al., Phys.~Rev.~Lett.~\textbf{98}, 263601 (2007).
\bibitem{morsch} H.~Lignier, et al., Phys.~Rev.~Lett.~\textbf{99}, 220403 (2007).
\bibitem{Milena_Grifoni} M.~Grifoni, P.~H\"anggi, Phys.~Rep.~\textbf{304}, 229 (1998).
\bibitem{acceleration_of_tunneling} V.~Averbukh, S.~Osovski, and N.~Moiseyev, Phys.~Rev.~Lett.~\textbf{89}, 253201 (2002).
\bibitem{imaginary_potential} V.~I.~Balykin and V.~S.~Letokhov, Atom Optics with Laser Light (Harwood Academic, Chur, Switzerland, 1995),
experimentally e.g. R.~St\"{u}tzle, et al.,
Phys.~Rev.~Lett.~\textbf{95}, 110405 (2005).
\bibitem{chaos_assisted_tunneling} D.~A.~Steck, W.~H.~Oskay, M.~G.~Raizen, Science \textbf{293}, 274 (2001).
W.~Hensinger et al., Nature \textbf{412}, 52 (2001).
\bibitem{weis} A.~Eckardt, C.~Weiss, and M.~Holthaus, Phys.~Rev.~Lett.~\textbf{95}, 260404 (2005). N.~Teichmann and C. Weiss,
EPL \textbf{78}, 1009 (2007).
\bibitem{Rachets_papers} e.g. H.~Schanz, M.F.~Otto, R.~Ketzmerick, and T.~Dittrich, Phys.~Rev.~Lett.~\textbf{87}, 070601
(2001).
\end{thebibliography}
\end{document}